\documentclass[aip,reprint
]{revtex4-2}

\usepackage{graphicx}
\usepackage{dcolumn}
\usepackage{amsmath,amssymb,bm}
\usepackage[mathlines]{lineno}
\usepackage[utf8]{inputenc}
\usepackage[T1]{fontenc}
\usepackage{mathptmx}
\usepackage{etoolbox}
\usepackage{extarrows}  
\usepackage[sort&compress]{natbib}
\usepackage[hidelinks]{hyperref}
\usepackage{xcolor}
\usepackage{ulem}

\newcommand{\ii}{\mathrm{i}}
\newcommand{\Rey}{\mathrm{Re}}
\newcommand{\Ma}{\mathrm{Ma}}
\newcommand{\Real}{\mathsf{Re}}

\makeatletter
\def\@email#1#2{%
\endgroup
\patchcmd{\titleblock@produce}
{\frontmatter@RRAPformat}
{\frontmatter@RRAPformat{\produce@RRAP{*#1\href{mailto:#2}{#2}}}\frontmatter@RRAPformat}
{}{}
}%
\makeatother

\begin{document}

\title{Closed-Form Solution for Oscillatory Flow and Wall Shear Stress in Axisymmetric Corrugated Tubes}

\author{F. Keil}
\affiliation{Helmholtz Institute Erlangen-N\"urnberg for Renewable Energy (IET-2), Forschungszentrum J\"ulich, 91058 Erlangen, Germany}
\affiliation{Department of Chemical and Biological Engineering, Friedrich-Alexander-Universit\"at Erlangen-N\"urnberg, 91058 Erlangen, Germany}

\author{P. Malgaretti}
\affiliation{Helmholtz Institute Erlangen-N\"urnberg for Renewable Energy (IET-2), Forschungszentrum J\"ulich, 91058 Erlangen, Germany}
\affiliation{Department of Physics, Friedrich-Alexander-Universit\"at Erlangen-N\"urnberg, 91058 Erlangen, Germany}
\email{p.malgaretti@fz-juelich.de}

\author{O. Aouane}
\affiliation{Helmholtz Institute Erlangen-N\"urnberg for Renewable Energy (IET-2), Forschungszentrum J\"ulich, 91058 Erlangen, Germany}
\affiliation{Department of Chemical and Biological Engineering, Friedrich-Alexander-Universit\"at Erlangen-N\"urnberg, 91058 Erlangen, Germany}
\email{o.aouane@fz-juelich.de}

\author{J. Harting}
\affiliation{Helmholtz Institute Erlangen-N\"urnberg for Renewable Energy (IET-2), Forschungszentrum J\"ulich, 91058 Erlangen, Germany}
\affiliation{Department of Chemical and Biological Engineering, Friedrich-Alexander-Universit\"at Erlangen-N\"urnberg, 91058 Erlangen, Germany}
\affiliation{Department of Physics, Friedrich-Alexander-Universit\"at Erlangen-N\"urnberg, 91058 Erlangen, Germany}

\date{\today}

\begin{abstract}
Pulsatile flow in corrugated tubes arises in hemodynamics and microfluidics, where oscillatory forcing and geometric constrictions jointly determine transport and wall loading. We extend Womersley's classical solution for oscillatory flow in a rigid circular tube to rigid axisymmetric tubes of slowly varying radius. Within the lubrication approximation, we derive closed-form expressions for the axial velocity profile, volumetric flow rate, phase lag, and wall shear stress at an arbitrary Womersley number. Three-dimensional lattice Boltzmann simulations are used to assess the regime of validity of the theory. The analysis shows that the local velocity profile varies strongly along the tube, ranging from plug-like in wide sections to more parabolic near bottlenecks. The cycle-maximum flow rate decreases with increasing corrugation, but this reduction weakens as pulsatility increases, reflecting a crossover from the quasi-steady scaling $\langle R^{-4}\rangle^{-1}$ to the high-frequency scaling $\langle R^{-2}\rangle^{-1}$. The wall shear stress is maximal at the bottleneck and decreases with Womersley number. For sinusoidal corrugations, the bottleneck wall shear stress depends non-monotonically on the corrugation because of the competition between local shear amplification and global hydraulic resistance. Closed-form expressions for the time-averaged wall shear stress and oscillatory shear index further connect the theory to standard hemodynamic metrics.
\end{abstract}

\maketitle

% ======================================================================

\section{Introduction}
Pulsatile flow is a key fluid-dynamic phenomenon. A prominent example is blood flow in arteries, driven by the heart's rhythmic pumping~\cite {zamir2016}. Beyond the cardiovascular system, the aqueous humor in the eye oscillates, as it is coupled to the cardiac cycle~\cite{johnstonea2011}, and cerebrospinal fluid within the cerebral ventricles likewise exhibits oscillatory flow patterns~\cite{linninger2016}. Pulsatile flow also plays an important role in technical applications: microfluidic devices exploit oscillatory flow to facilitate droplet formation, eliminating the dependence on the Rayleigh--Plateau instability, to enhance mixing under laminar conditions where transport is otherwise diffusion-limited, and to separate cells based on their size and deformability with a funnel ratchet~\cite{dincau2019}. More recently, oscillatory flow has been shown to enable inertial focusing of sub-micrometer particles in short microchannels, circumventing the need for long channel distances required by steady flow~\cite{mutlu2018}. For recent broad reviews of pulsatile flow modeling, numerical methods, and applications in biomedical and engineering systems, see, e.g., Ref.~\cite{Roknujjaman2026}.

\begin{figure}
\centering
\includegraphics[width=.85\linewidth]{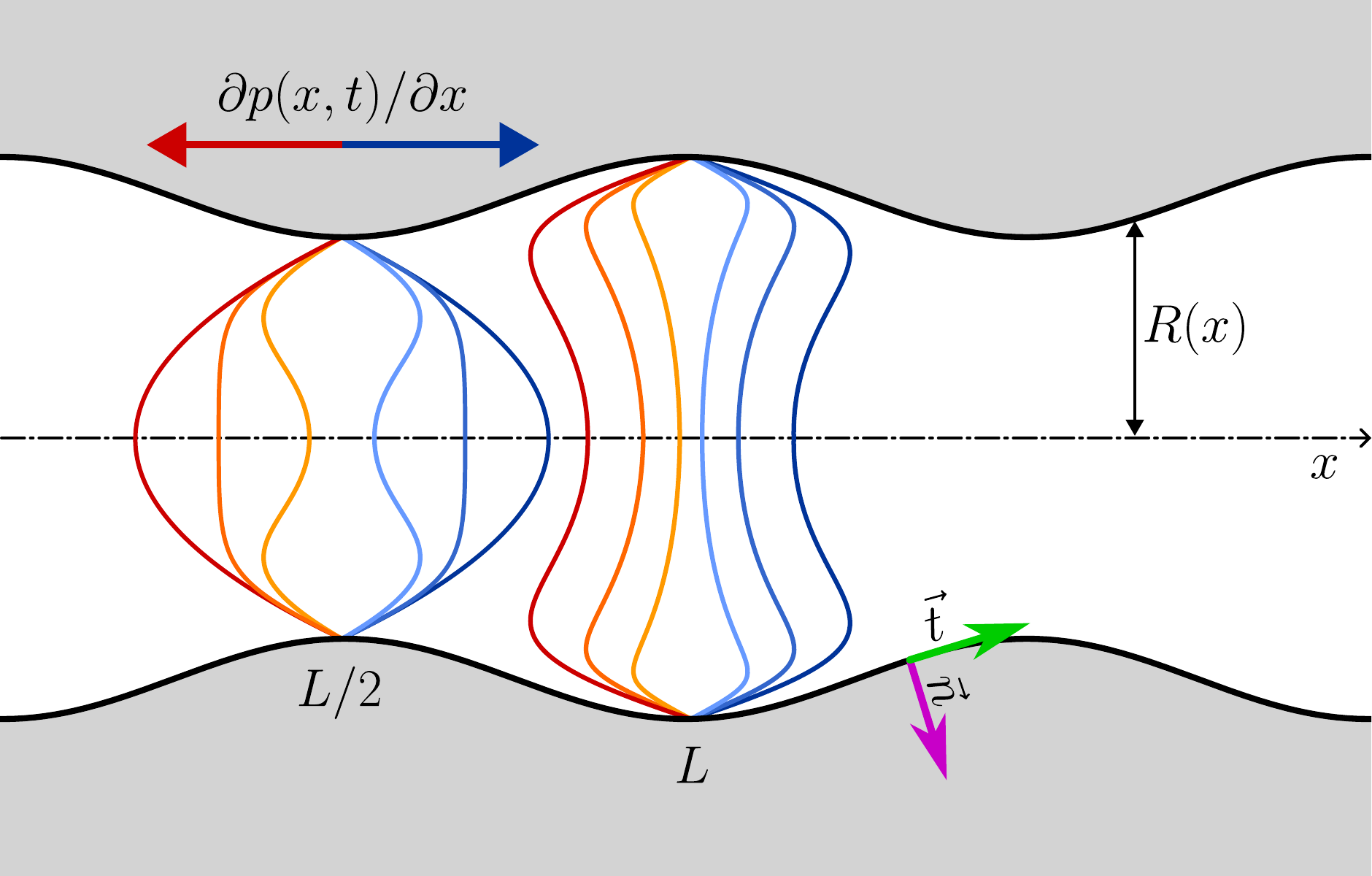}
\caption{Schematic drawing of a pulsatile flow in a corrugated tube of radius $R(x)$ driven by a time-dependent axial pressure gradient $\partial p(x,t)/\partial x$. The unit normal vector $\vec{n}$ and the unit tangential vector $\vec{t}$ at the tube wall are indicated. $L$ denotes the length of the tube.}
\label{fig:schematic}
\end{figure}

While pulsatility governs the flow's temporal characteristics, spatial features such as geometric constrictions modulate its local behavior. For instance, stenosis, the pathological narrowing of blood vessels, alters the hemodynamic environment and threatens the cardiovascular system~\cite{lusis2000,berger2000}. Hemodynamic wall shear stress is closely related to the localization and progression of atherosclerotic plaques: low wall shear stress promotes plaque development, whereas high wall shear stress is linked to plaque destabilization and rupture~\cite{malek1999,samady2011,thim2012,song2022,mazzi2024}. Conversely, the altered flow dynamics of pulsatile flow in corrugated tubes have led to a novel measurement technique for quantifying cell--cell adhesion in circulating tumor cell clusters, which are more potent initiators of metastasis than individual circulating tumor cells~\cite{bottos2014,mutlu2020}. Additionally, drug carriers for the treatment of stenosis and contrast agents used for diagnosis must adhere to the vessel wall at the stenotic region, and the local wall shear stress is one of the main factors determining adhesion~\cite{forouzandehmehr2022,decuzzi2006,zhou2023,pabi2026}.

Analytical descriptions of pulsatile flow date back to Sexl and Womersley, who derived the velocity profile for an incompressible, Newtonian fluid oscillating in a rigid, straight, circular tube~\cite{sexl1930,womersley1955rigid}. Subsequent extensions consider elastic and rigid curved tubes with circular cross sections, as well as rigid curved tubes with arbitrary cross sections~\cite{womersley1955elasticI,womersley1958elasticII,lyne1971,smith1975}. Rao and Devanathan~\cite{ramachandradevanathan1973} studied pulsatile flow in tubes of varying cross section at low Reynolds number and derived wall shear stress distributions for tapered, locally constricted, and peristaltic geometries. Their analysis, however, is restricted to the quasi-steady (low-frequency) regime and does not retain the full Bessel-function structure of Womersley's solution at arbitrary frequency. Rao~\cite{ramachandrarao1983} subsequently considered oscillatory flow in an elastic, corrugated tube under the lubrication approximation, deriving a second-order ordinary differential equation (ODE) for the excess pressure that drives the flow.
However, no closed-form expression of the wall shear stress was obtained. In the rigid-wall limit, Rao's ODE reduces to an algebraic relation for the pressure gradient~\cite{zhang2024}, which is the starting point of the present work. More recently, Pande \textit{et al.}~\cite{pande2023,huang2025,pandeboyko2026} developed reduced-order theories for oscillatory flow in compliant conduits at arbitrary Womersley number under the lubrication approximation, extended them to three-dimensional deformable micro-channels, and used the Lorentz reciprocal theorem to derive corrections to the steady flow rate due to slow oscillations in rigid two-dimensional channels of nonuniform width. The interplay between wall compliance and longitudinal deformation was analyzed for peristaltic flows~\cite{shapiro1969,takagi2011,winn2026}. For non-Newtonian fluids, pulsatile flow has been addressed between parallel plates~\cite{fan2026} and within three-dimensional deformable, axisymmetric tubes~\cite{roldan2026} using the Oldroyd-B model. Kannaiyan \textit{et al.}~\cite{kannaiyan2026} recently derived solutions for the sudden cessation of pulsatile flow in rigid, straight circular tubes. Besides pure fluid flows, transport through corrugated channels~\cite{kitanidis1997,mangeat2018,malgaretti2013,chassagne2019} has been studied under steady-flow conditions, where higher-order corrections to the lubrication theory have been developed~\cite{tavakol2017}. In a complementary vein, the lattice Boltzmann method has become a powerful tool for simulating pulsatile flows in complex vascular geometries, including stenosed and stented arteries~\cite{wang2020,osaki2021,shi2014,peruzzo2024,ding2021,dhawan2025,rajmane2025,neflas2026}. The lattice Boltzmann method has also been used to study cerebrospinal fluid flow within stenosed brain ventricles~\cite{koh2021}.
Despite these advances, to the best of our knowledge, no closed-form description of the velocity field and wall shear stress for pulsatile flow in rigid corrugated tubes at arbitrary Womersley numbers has been reported. 

In this contribution, we extend Womersley's original solution to the case of a rigid, axisymmetric corrugated tube within the lubrication approximation. Unlike the elastic-tube theory of Rao~\cite{ramachandrarao1983}, the rigid-wall limit admits an algebraic (rather than differential) equation for the pressure gradient, which we exploit to obtain closed-form expressions for the axial velocity, the volumetric flow rate, the phase lag, and the wall shear stress. The resulting expressions are valid at arbitrary Womersley number, bridging the quasi-steady regime of Rao and Devanathan~\cite{ramachandradevanathan1973} and the compliant-conduit theory of Pande \textit{et al.}~\cite{pande2023}. Three-dimensional lattice Boltzmann simulations are used to identify the regime of validity of the lubrication approximation.

The remainder of the paper is organized as follows. Section~\ref{sec:theory} reviews Womersley's solution for pulsatile flow in straight, rigid tubes and presents the lubrication-based extension for axisymmetric corrugated tubes, including the derivation of the wall shear stress. Section~\ref{sec:numerics} describes the lattice Boltzmann method used to validate the analytical predictions and identifies the relevant length-scale ratios governing the lubrication approximation. Section~\ref{sec:results} presents the characteristic flow features, including the spatial heterogeneity of the velocity profile, the volumetric flow rate, the phase relationship between driving pressure and flow, and the wall shear stress. Finally, Section~\ref{sec:conclusion} summarizes the main findings.

%=====================================================================
\section{Analytical framework} \label{sec:theory}

In the following subsections, we recall Womersley's solution for pulsatile flow in straight, rigid tubes, Sec.~\ref{subsec:Womersley flow}, present the extension for corrugated tubes under the lubrication approximation, Sec.~\ref{subsec:Extension}, and derive the wall shear stress based on the velocity profile, Sec.~\ref{subsec:wall shear stress}.

\subsection{Womersley flow in a straight tube} \label{subsec:Womersley flow}

A Womersley flow is a pulsatile, laminar, axisymmetric flow of a homogeneous, incompressible, Newtonian fluid in a rigid circular tube of radius $R$. The Newtonian assumption is valid for blood under pulsatile, moderate-Reynolds-number conditions typical of larger arteries~\cite{samady2011,berger2000,shi2026}. The flow is driven by an axial oscillating pressure gradient,
\begin{equation}
\frac{\partial p(t)}{\partial x} = \Real \left\{ \mathcal{P} e^{\ii \omega t} \right\}\,,
\label{eq:oscillating pressure gradient straight}
\end{equation}
where $\mathcal{P}$ is the amplitude of the pressure gradient, $\omega$ is the angular frequency, $t$ is the time, $\ii$ is the imaginary unit, and $\Real \{\cdot\}$ denotes the real part. The axial velocity profile that satisfies the no-slip condition at the wall reads~\cite{womersley1955rigid},
\begin{equation}
u_x(r,t) = \Real \left\{ \frac{\ii \mathcal{P}}{\rho \omega}\left[1-\frac{J_{0}\!\left(r \sqrt{\frac{\omega}{\nu}} \ii^{3/2} \right)}{J_{0}\!\left(R \sqrt{\frac{\omega}{\nu}} \ii^{3/2}\right)}\right] e^{\ii \omega t}\right\}\,,
\label{eq:velocity straight}
\end{equation}
where $r$ is the radial coordinate, $\rho$ is the fluid density, $\nu$ is the kinematic viscosity, and $J_0$ denotes the zeroth-order Bessel function of the first kind. Note that the pressure gradient and the axial velocity carry opposite signs: a negative pressure gradient drives a positive (forward) velocity.

The dimensionless Womersley number,
\begin{equation}
\alpha = R \sqrt{\frac{\omega}{\nu}}\,,
\label{eq:womersley number}
\end{equation}
characterizes the velocity-profile shape. It can be rewritten in terms of Stokes viscous penetration depth~\cite{fung1984} $\delta$ as $\alpha = R/\delta$, with $\delta = \sqrt{\nu/\omega}$. When the oscillation period $T = 2\pi/\omega$ is much larger than the viscous diffusion time $\tau_\nu \sim R^2/\nu$, i.e.\ $\alpha \ll 1$, viscous forces dominate, and the flow has enough time to develop fully. It then resembles a quasi-steady Poiseuille profile. In the opposite limit, $\alpha \gtrsim 1$, transient inertia forces dominate, and a plug-like core with thin oscillatory boundary layers emerges.

\subsection{Extension to corrugated tubes} \label{subsec:Extension}

We extend Womersley's solution to rigid, axisymmetric tubes whose radius $R(x)$ varies slowly in the axial direction, i.e.\ under the lubrication approximation,
\begin{equation}
\left|\frac{\partial R(x)}{\partial x}\right| \ll 1\,,
\label{eq:lubrication}
\end{equation}
neglecting terms of order $O\!\bigl((\partial R/\partial x)^2\bigr)$ and higher. At leading order, the velocity profile retains the local Womersley form,
\begin{align}
u_x(x, r,t) = \Real \left\{ \frac{\ii \mathcal{P}(x)}{\rho \omega}\left[1-\frac{J_{0}\!\left(\alpha(x) \ii^{3/2} \frac{r}{R(x)}\right)}{J_{0}\!\left(\alpha(x) \ii^{3/2}\right)}\right] e^{\ii \omega t}\right\}\,,
\label{eq:velocity corrugated}
\end{align}
where $\mathcal{P}(x)$ is the local amplitude of the pressure gradient and $\alpha(x)$ is the local Womersley number, i.e., $\alpha(x) = R(x)\sqrt{\omega/\nu}$. Since the Stokes penetration depth $\delta$ is uniform along the tube, variations in the tube radius $R(x)$ directly modulate $\alpha(x)$, leading to spatially heterogeneous velocity profiles. For later reference, we define the Womersley number based on the mean tube radius $R_0$ as
\begin{equation}
\alpha_0 \equiv R_0\sqrt{\frac{\omega}{\nu}}\,.
\label{eq:mean womersley number}
\end{equation}

The local pressure gradient amplitude $\mathcal{P}(x)$ is determined by requiring that the volumetric flow rate
\begin{equation}
Q(t) = 2 \pi \int_0^{R(x)} u_x(x, r, t)\, r\, dr\,
\label{eq:flow rate}
\end{equation}
is homogeneous in $x$, i.e.\ $\partial Q/\partial x = 0$. Carrying out the integration and imposing this constraint, it is convenient to introduce the complex effective conductance of a cross section,
\begin{equation}
\mathcal{D}(x) \equiv R^2(x) \left[ \frac{1}{2} - \frac{J_1 \!\left( \alpha(x) \ii^{3/2} \right)}{\alpha(x) \ii^{3/2} J_0\!\left( \alpha(x) \ii^{3/2} \right)} \right],
\label{eq:conductance}
\end{equation}
which absorbs the Bessel-function structure. In the quasi-steady limit $\alpha(x) \ll 1$, $\mathcal{D}(x) \sim \ii\omega R^4(x)/(16\nu)$ (Appendix~\ref{app:conductance_asymptotics}), so that Eq.~\eqref{eq:flow rate corrugated} reduces to the Hagen--Poiseuille law for a corrugated tube. The local pressure gradient amplitude then reads
\begin{equation}
\mathcal{P}(x) = \frac{\Delta P}{\mathcal{D}(x)\displaystyle\int_0^L \frac{dx'}{\mathcal{D}(x')}}\,,
\label{eq:local pressure gradient}
\end{equation}
where $\Delta P = P_2 - P_1$ is the imposed pressure difference across a tube of length $L$. For forward flow from $x=0$ to $x=L$, one has $P_1>P_2$ and therefore $\Delta P<0$.
The volumetric flow rate is
\begin{equation}
Q(t) = \Real \left\{ \frac{2 \pi \ii  \Delta P}{\rho \omega}\, e^{\ii \omega t}  \left( \int_0^L \frac{dx}{\mathcal{D}(x)} \right)^{\!-1}\right\}\,.
\label{eq:flow rate corrugated}
\end{equation}
Compared with the local oscillating pressure gradient, $\partial p/\partial x = \Real \{\mathcal{P}(x)\,e^{\ii\omega t}\}$, the flow rate contains additional imaginary factors beyond $e^{\ii\omega t}$, which produce a phase lag $\Delta\phi$ between the pressure forcing and the flow response. Collecting these factors into a complex amplitude
\begin{equation}
\chi  
= \frac{2 \pi \ii  \Delta P}{\rho \omega} \left( \int_0^L \frac{dx}{\mathcal{D}(x)} \right)^{\!-1}\,,
\label{eq:chi}
\end{equation}
the phase lag is given by
\begin{equation}
\Delta \phi = \arg(\chi)\,.
\label{eq:phase lag}
\end{equation}
In the uniform-radius limit, $R(x)=R$ and $\alpha = \text{const.}$, Eqs.~\eqref{eq:velocity corrugated}, \eqref{eq:flow rate corrugated}, and \eqref{eq:chi} reduce to the classical Womersley solutions.

\subsection{Wall shear stress} \label{subsec:wall shear stress}

The wall shear stress (WSS) is defined as the tangential component of the viscous traction at the tube wall,
\begin{equation}
\tau_\parallel = \vec{n} \cdot \boldsymbol{\tau} \cdot \vec{t}\,,
\label{eq:wss definition}
\end{equation}
where $\boldsymbol{\tau} = \eta \bigl( \nabla \vec{u} + (\nabla \vec{u})^\intercal \bigr)$ is the viscous stress tensor, $\eta=\rho \nu$ is the dynamic viscosity, and $\vec{n}$ and $\vec{t}$ are the outward unit normal and the unit tangent vectors at the wall, respectively (Fig.~\ref{fig:schematic}). The lubrication approximation allows us to use the small-angle approximation, such that the full expression reads
\begin{equation}
\tau_\parallel = \eta \left( -2 \frac{\partial u_r}{\partial r} \frac{\partial R}{\partial x}  - \frac{\partial u_x}{\partial r} + 2\frac{\partial u_x}{\partial x} \frac{\partial R}{\partial x} \right)\,,
\label{eq:wss full}
\end{equation}
where each velocity derivative is evaluated at $r = R(x)$.
The key simplification arises from the structure of the Bessel-function argument in the numerator of Eq.~\eqref{eq:velocity corrugated}: $\alpha(x)\,\ii^{3/2}\, r/R(x) = r\,\ii^{3/2}\sqrt{\omega/\nu}$, which is independent of $x$. Differentiating Eq.~\eqref{eq:velocity corrugated} with respect to $x$ and evaluating at the wall, the inner Bessel function $J_0(r\,\ii^{3/2}\sqrt{\omega/\nu})$ contributes no $x$-derivative, and the terms proportional to \(\partial_x \mathcal{P}\) drop out because the bracket multiplying \(\mathcal{P}(x)\) in Eq.~\eqref{eq:velocity corrugated} vanishes identically at \(r = R(x)\). One finds
\begin{equation}
\left.\frac{\partial u_x}{\partial x}\right|_{r=R(x)} = -\left.\frac{\partial u_x}{\partial r}\right|_{r=R(x)} \cdot \frac{\partial R}{\partial x}\,,
\label{eq:dux_dx identity}
\end{equation}
such that the third term in Eq.~\eqref{eq:wss full} is of order $O\!\bigl((\partial R / \partial x)^2)$ and negligible.
Similarly, axisymmetric incompressibility, $\partial_x u_x + r^{-1}\partial_r(r u_r) = 0$, evaluated at the wall where the no-penetration condition $\vec u\cdot\vec n = 0$ implies $u_r|_{r=R(x)} = O(\partial R/\partial x)$, gives
$\partial u_r/\partial r\big|_{r=R(x)} = -\partial u_x/\partial x\big|_{r=R(x)}$
at leading order in $\partial R/\partial x$ (see Appendix~\ref{app:partial_derivatives}). Both derivatives at $r=R(x)$ are therefore $O(\partial R/\partial x)$, so the product $(\partial u_r/\partial r)(\partial R/\partial x)$ is $O\!\bigl((\partial R/\partial x)^2\bigr)$ and negligible. The wall shear stress, therefore, simplifies to
\begin{equation}
\tau_\parallel = - \eta\, \Real \left\{ \frac{\mathcal{P}(x)\, \ii^{5/2}}{\rho \omega} \sqrt{\frac{\omega}{\nu}}\; \frac{J_1\!\left( \alpha(x)\, \ii^{3/2} \right)}{J_0\!\left( \alpha(x)\, \ii^{3/2} \right)}\; e^{\ii \omega t} \right\}\,.
\label{eq:wss final}
\end{equation}

The same identities yield the normal component of the viscous traction at the wall. Using $\tau_\perp = \vec{n} \cdot \boldsymbol{\tau} \cdot \vec{n} = 2\eta\bigl(\partial u_r/\partial r - (\partial u_x/\partial r)(\partial R/\partial x)\bigr)$ and substituting $\partial u_r/\partial r\big|_{r=R(x)} = -\partial u_x/\partial x\big|_{r=R(x)}$ at leading order together with Eq.~\eqref{eq:dux_dx identity}, one obtains $\tau_\perp = 0$ at leading order. This shows that for an incompressible Newtonian fluid in an axisymmetric tube of slowly varying radius, the normal viscous stress vanishes at the wall and only the tangential stress, Eq.~\eqref{eq:wss final}, survives. The total normal stress, of course, includes the pressure and is therefore nonzero. This result has practical significance for bio-adhesion modeling~\cite{decuzzi2006} because the mechanical loading on the endothelium or on adhered particles is entirely tangential at leading order, so that adhesion models only need to incorporate $\tau_\parallel$ rather than the full traction vector.

% ======================================================================
\section{Numerical validation} \label{sec:numerics}

To identify the regime in which the lubrication approximation is valid, we compare the analytical predictions with three-dimensional lattice Boltzmann (LBM) simulations.

\subsection{Lattice Boltzmann method}

The flow is resolved on a D3Q19 lattice with speed of sound $c_s = 1/\sqrt{3}$. A single-relaxation-time Bhatnagar--Gross--Krook (BGK) collision operator is used with a relaxation time of one, yielding a kinematic viscosity $\nu = 1/6$ in lattice units~\cite{krueger2017}. The oscillating pressure gradient is mimicked by a time-dependent body force $f_b(t) = f_0 \sin(\omega t)$, which is incorporated using the Guo forcing scheme~\cite{guo2002}. At the tube wall, the no-slip condition is enforced via the bounce-back scheme, which has been shown to predict wall shear stress reliably in complex vascular geometries~\cite{osaki2021}. Periodic boundary conditions are applied at the inlet and outlet, exploiting flow symmetry at low Reynolds number.

The Reynolds number is defined as $\Rey = 2 R_{\min} u_\text{Poiseuille}^{\max}/\nu$, where $u_\text{Poiseuille}^{\max}$ is the centerline velocity of the corresponding steady Poiseuille flow evaluated at the bottleneck, and the Mach number as $\Ma = u_\text{Poiseuille}^{\max}/c_s$. All simulations satisfy $\Ma < 0.1$, thereby recovering the incompressible Navier--Stokes equations~\cite{krueger2017}. The simulation setup was validated against the analytical Womersley solution in a straight tube for Womersley numbers ranging from $\alpha = 1$ to $\alpha = 13$, with relative errors below $3\,\%$.

\subsection{Corrugated tube geometry and lubrication assessment}

We choose a corrugated tube with sinusoidally varying radius,
\begin{equation}
R(x) = R_0 + R_1 \cos\!\left( \frac{2\pi x}{L} \right),
\label{eq:tube geometry}
\end{equation}
as a representative geometry modeling stenotic vessels (Fig.~\ref{fig:schematic}). Here, $R_0$ is the mean radius, $R_1$ is the modulation amplitude, and $L$ is the tube length. The specific parameters used are $R_0 = 30$, $R_1 = 3$, and $L = 352$ in lattice units, giving a relative amplitude $\varepsilon = R_1/R_0 = 0.10$ and a slope ratio $R_1/L \approx 0.009 \ll 1$ that satisfies the lubrication condition [Eq.~\eqref{eq:lubrication}].

To assess the validity of the lubrication approximation systematically, we compared analytical and numerical steady Poiseuille profiles for a series of corrugated tubes with slope ratios $R_1/L$ ranging from $0.094$ to $0.009$ (Table~\ref{tab:tube geometries}). The relative error decreases systematically with diminishing slope and follows an approximately quadratic scaling $\xi \propto (R_1/L)^2$, consistent with the expected leading-order correction to the lubrication theory~\cite{tavakol2017}. For the geometry used in this work ($R_1/L = 0.009$), the error is below $3\,\%$, confirming that the lubrication framework is appropriate.

An additional constraint specific to the lattice-Boltzmann method arises from the finite lattice speed of sound: the pressure relaxation time $T_p = L/c_s$ must remain much smaller than the oscillation period $T$ for the flow to behave incompressibly within the simulation~\cite{krueger2017}. In physical systems such as blood, the speed of sound in the fluid is $\sim 1500\;\text{m/s}$, such that this constraint is trivially satisfied. We restrict the average Womersley number to $\alpha_0 \leq 4$, corresponding to $T_p/T \leq 0.288$, for which the flow-rate deviation between $x = 0$ and $x = L/2$ remains below $\sim 5\,\%$.

\begin{table}[h!]
\caption{Corrugated-tube geometries used for the lubrication-approximation validation (in lattice units). The mean radius $R_0 = 30$ is held constant; the ratio $R_1/L$ decreases from left to right.}
\begin{tabular}{l c c c c c}
\hline\hline
$R_1$ & 9 &  6 & 6 & 3 & 3 \\
$L$ & 96 & 96 & 128 & 128 & 352 \\
$R_1/L$ & 0.094 & 0.063 & 0.047 & 0.023 & 0.009 \\
\hline\hline
\end{tabular}
\label{tab:tube geometries}
\end{table}

\subsection{Comparison of analytical and numerical velocity profiles}
The agreement between the analytical $u_a$ and numerical $u_s$ velocity profiles, evaluated on a two-dimensional slice containing the centerline of the pipe, is quantified by the relative error,
\begin{equation}
\xi = \frac{\displaystyle\sum_x \sum_t \sum_r \left|u_a(x, r, t) - u_s(x, r, t)\right|}{\displaystyle\sum_x \sum_t \sum_r \left|u_a(x, r, t)\right|}\,,
\label{eq:relative_error}
\end{equation}
where the summation runs over all sampled radial positions $r$, time points $t$ within one period, and axial positions $x$ along the tube. Owing to axisymmetry, the velocity field is independent of the azimuthal coordinate, and the two-dimensional representation is equivalent to the full three-dimensional solution. 

The radial summation is implemented as a simple unweighted sum over the discrete grid points; that is, all sampled radial positions contribute equally to the error metric. Accordingly, Eq.~\eqref{eq:relative_error} corresponds to a discrete $\mathrm{L}^1$ norm evaluated on a uniform grid in $r$, $t$, and $x$.

\begin{figure}[h!]
\centering
\includegraphics[width=\linewidth]{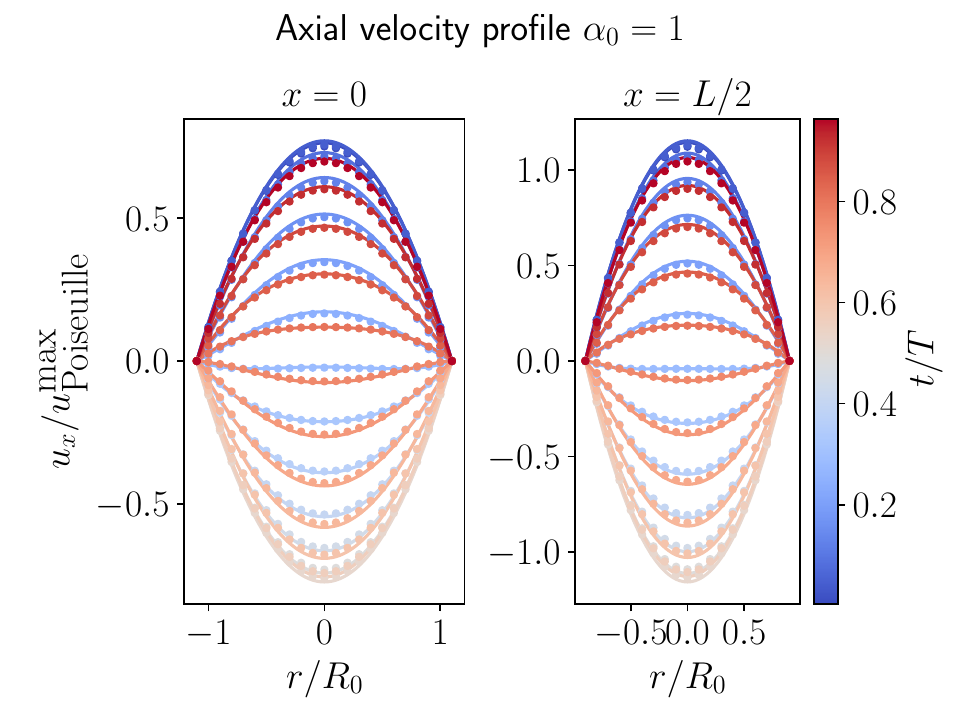}
\caption{Axial velocity profiles at $x = 0$ (widest section, left) and $x = L/2$ (bottleneck, right) for $\alpha_0 = 1$: LBM results (symbols) vs.\ analytical prediction (lines). The velocity $u_x$ is normalized by the centerline Poiseuille velocity $u_\text{Poiseuille}^{\max}$ and the radial coordinate by the mean radius $R_0$. Line colors indicate the phase $t/T$ within the oscillation cycle. The relative error is $\xi < 3.5\,\%$.}
\label{fig:alpha1}
\end{figure}

\begin{figure}[h!]
\centering
\includegraphics[width=\linewidth]{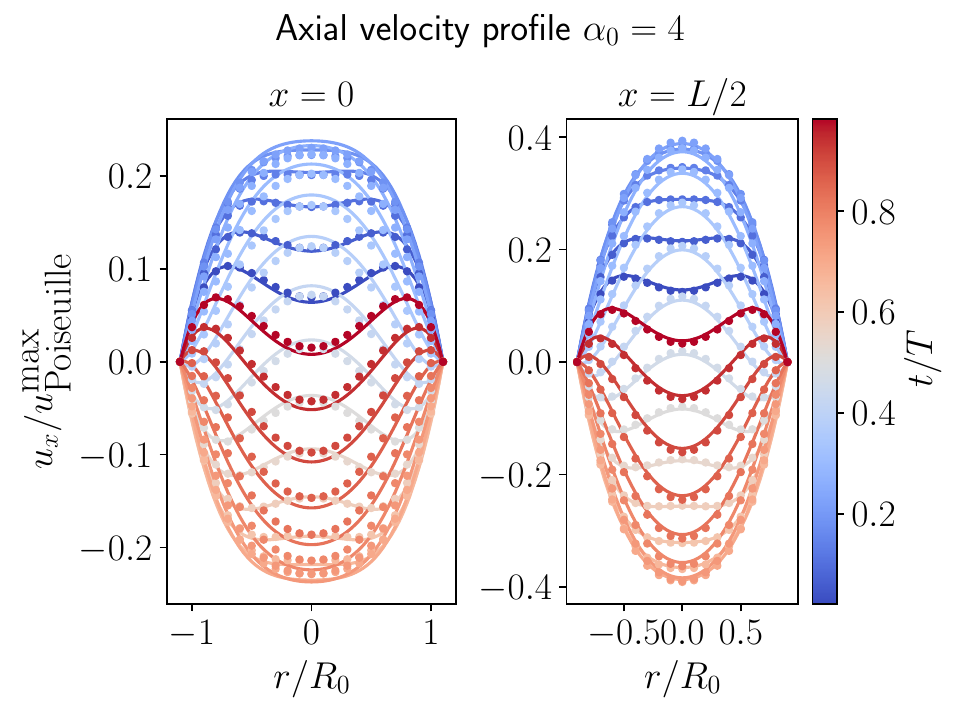}
\caption{Same as Fig.~\ref{fig:alpha1} but for $\alpha_0 = 4$. At the widest section, the profile is more plug-like, while at the bottleneck it remains close to parabolic. The relative error is $\xi < 3.5\,\%$.}
\label{fig:alpha4}
\end{figure}

Figures~\ref{fig:alpha1} and \ref{fig:alpha4} compare the velocity profiles for $\alpha_0 = 1$ and $\alpha_0 = 4$, respectively, at $x = 0$ (widest section) and $x = L/2$ (bottleneck). In all cases, $\xi$ remains below $3.5\,\%$, confirming the validity of the lubrication-based analytical framework for the geometry under consideration.

% ======================================================================
\section{Results} \label{sec:results}

Having validated the analytical framework, we now examine the characteristic features of pulsatile flow in corrugated tubes.

\subsection{Spatial heterogeneity of the velocity profile}

\begin{figure}[h!]
\centering
\includegraphics[width=\linewidth]{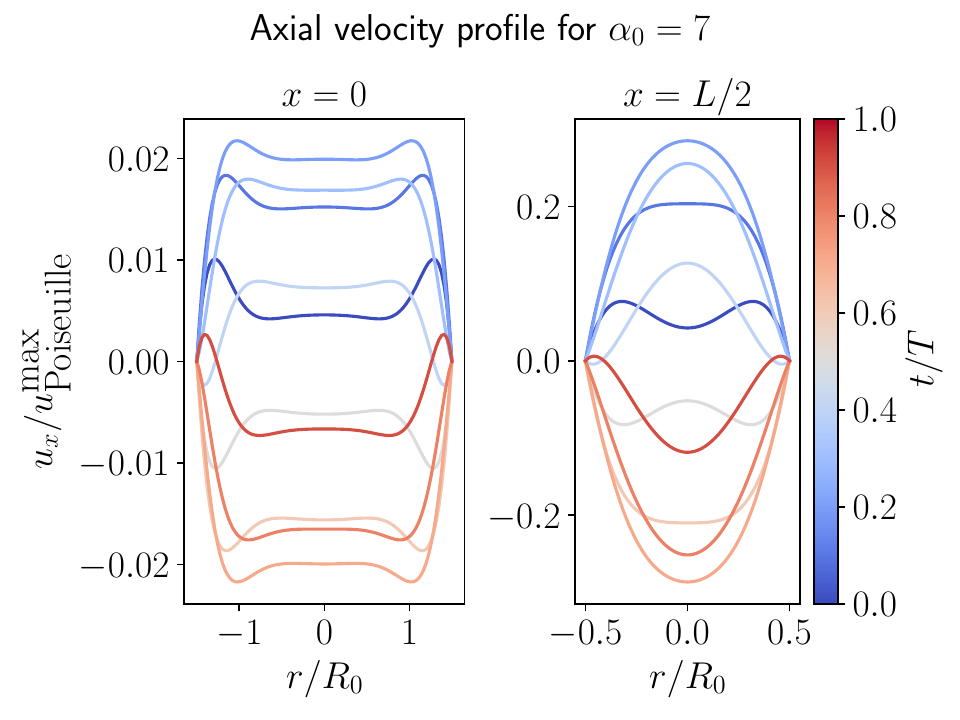}
\caption{Spatial heterogeneity of the axial velocity profile for $\alpha_0 = 7$ and $R_1/R_0 = 0.5$ ($\Delta S \approx 2.2$) at several phases $t/T$. Left: widest section ($x = 0$, local $\alpha = 10.5$), showing a plug-like core. Right: bottleneck ($x = L/2$, local $\alpha = 3.5$), showing a nearly parabolic profile. Velocities are normalized by $u_\text{Poiseuille}^{\max}$ and radial positions by $R_0$.}
\label{fig:spatial heterogeneity}
\end{figure}

In contrast to steady Poiseuille flow in a corrugated tube, where the velocity profile remains parabolic along the tube and only changes amplitude with tube radius, pulsatile flow exhibits spatial heterogeneity in both amplitude and shape. Figure~\ref{fig:spatial heterogeneity} illustrates this by comparing the axial velocity at $x=0$ (widest section) and $x=L/2$ (bottleneck) for a tube with $\alpha_0 = 7$ at several phases within the cycle.

The physical origin of this heterogeneity is the position-dependent Womersley number $\alpha(x)$. Because the Stokes penetration depth $\delta$ is uniform along the tube, a larger local radius corresponds to a larger local $\alpha$. At the widest section, $\alpha(0) = 10.5$, so the viscous boundary layer occupies only a thin annulus near the wall, and the core of the flow moves as a nearly rigid plug. At the bottleneck, $\alpha(L/2) = 3.5$, the boundary layer extends further toward the centerline, recovering a more parabolic shape. The velocity amplitude is also strongly redistributed along the tube: for the parameters shown, the amplitude at the bottleneck is roughly an order of magnitude larger than at the widest section, reflecting the local pressure-gradient amplification $\mathcal{P}(x)$ required to maintain a uniform flow rate. The ratio $u_{x,\max}(R_{\min})/u_{x,\max}(R_{\max}) = R_{\max}^2/R_{\min}^2 = 9$ predicted in Appendix~\ref{app:ratio-v} for the quasi-steady limit is closely reproduced by the data in Fig.~\ref{fig:spatial heterogeneity} even at $\alpha_0 = 7$, because both the quasi-steady and high-frequency limits yield $u_{x,\max}\propto 1/R^2(x)$. The wall shear stress depends both on the local velocity gradient and on the local pressure gradient amplitude $\mathcal{P}(x)$; their interplay is quantified in Sec.~\ref{sec:wss}.

Two limiting regimes clarify these observations. In the quasi-steady limit $\alpha \to 0$ in a straight tube, the Bessel-function ratio in Eq.~\eqref{eq:velocity straight} reduces to the parabolic profile $1 - r^2/R^2(x)$, and the local velocity amplitude scales as $R^2(x)$. In the high-frequency limit $\alpha_0 \gg 1$ in a straight tube, the profile develops a flat core with a thin oscillatory boundary layer of thickness $\delta$. The corrugated tube thus transitions continuously between these limits as a function of position: the widest section is in the high-$\alpha$ regime, while the bottleneck remains in an intermediate regime where viscous diffusion still shapes the profile.

\subsection{Volumetric flow rate and phase lag}

\begin{figure}[h!]
\centering
\includegraphics[width=\linewidth]{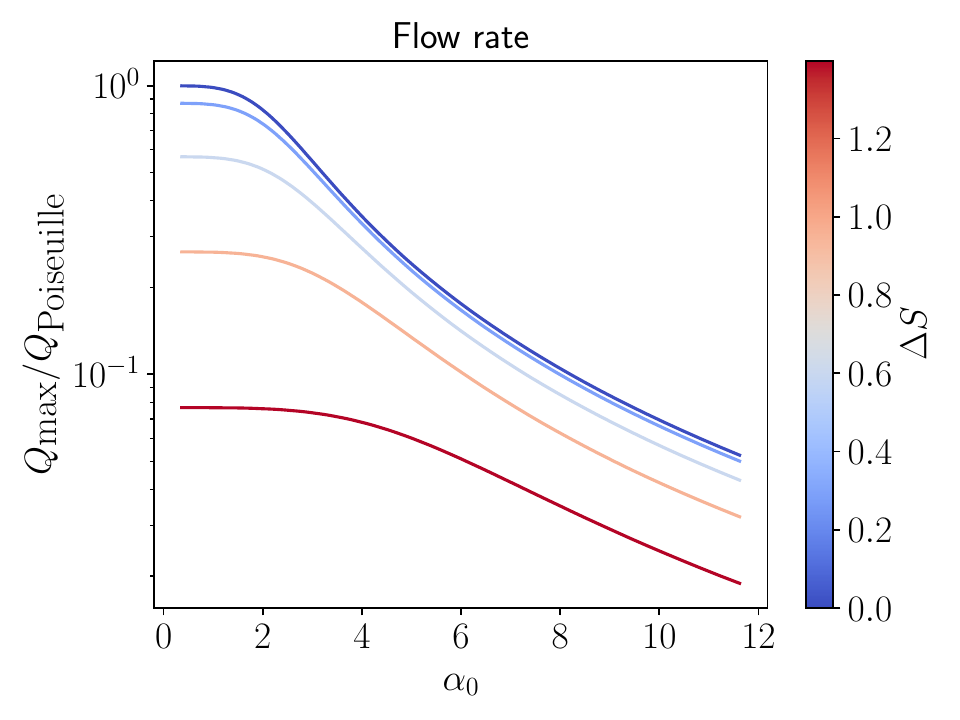}
\caption{Cycle-maximum volumetric flow rate $Q_{\max}/Q_\text{Poiseuille}$ vs.\ average Womersley number $\alpha_0$ for several degrees of corrugation $\Delta S$ [Eq.~\eqref{eq:corrugation}]. The corrugation acts as an additional resistance that is most pronounced at low $\alpha_0$.}
\label{fig:flow rate}
\end{figure}

Because the net volumetric flow rate vanishes for a purely sinusoidal pressure gradient, we characterize the flow by the cycle-maximum volumetric flow rate $Q_{\max}$, normalized by the steady Poiseuille flow rate $Q_\text{Poiseuille} = \pi R_0^4 |\Delta P| / (8 \eta L)$ in a straight tube of radius $R_0$. Figure~\ref{fig:flow rate} shows $Q_{\max}/Q_\text{Poiseuille}$ as a function of the average Womersley number $\alpha_0$ for several degrees of corrugation,
\begin{equation}
\Delta S = 2 \ln \!\left( \frac{R_0 + R_1}{R_0 - R_1} \right).
\label{eq:corrugation}
\end{equation}

Two trends are apparent. First, $Q_{\max}$ decreases monotonically with increasing $\alpha_0$: at higher pulsatility, the velocity profile has less time to develop within each half-cycle, reducing the peak flow. Second, $Q_{\max}$ decreases with increasing $\Delta S$ at fixed $\alpha_0$, because the constriction acts as a hydraulic resistance. Importantly, the difference between the straight-tube and the corrugated-tube flow rates is larger for small $\alpha_0$ (quasi-steady limit) than for large $\alpha_0$. This can be understood from the fact that, in the quasi-steady limit, the flow rate scales as $R^4$ (Hagen--Poiseuille law), so the constriction has a strong effect. At high pulsatility, the plug-like profile reduces sensitivity to tube radius, and corrugation has a comparatively smaller impact.

The asymptotic behavior of Eq.~\eqref{eq:flow rate corrugated} in the two limits further supports this picture. For $\alpha_0 \ll 1$, expanding the Bessel functions yields $Q_{\max} \propto \langle R^{-4} \rangle^{-1}$, where $\langle \cdot \rangle$ denotes the spatial average over one period (Appendix~\ref{app:conductance_asymptotics}). The harmonic mean of $R^{-4}$ is dominated by the bottleneck, making the flow rate highly sensitive to the corrugation amplitude. For $\alpha_0 \gg 1$, the flow rate depends on $\langle R^{-2} \rangle^{-1}$, a weaker dependence on the local radius. The crossover between these two scalings occurs for $\alpha_0 = O(1)$ and is consistent with the convergence of the curves in Fig.~\ref{fig:flow rate} at large $\alpha_0$.

\begin{figure}[h!]
\centering
\includegraphics[width=\linewidth]{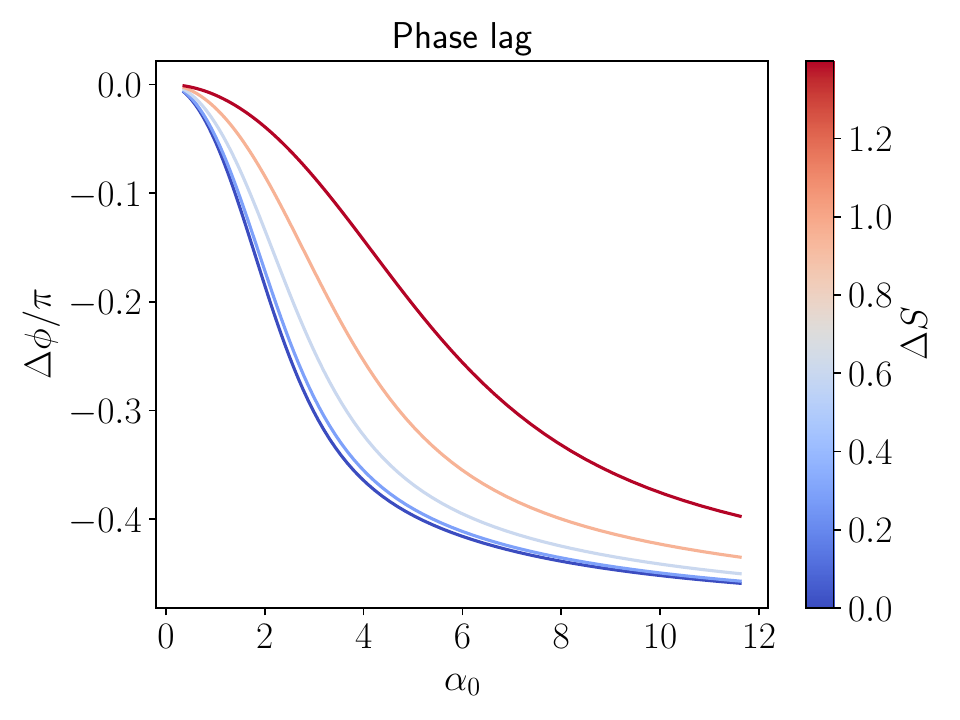}
\caption{Phase lag $\Delta\phi/\pi$ between the oscillatory pressure gradient and the volumetric flow rate vs.\ $\alpha_0$ for several degrees of corrugation $\Delta S$. Higher corrugation reduces $|\Delta\phi|$ at a given $\alpha_0$.}
\label{fig:phase lag}
\end{figure}

The complex amplitude $\chi$ [Eq.~\eqref{eq:chi}] that determines $Q_{\max}$ also encodes the phase lag $\Delta\phi$ between the oscillatory pressure gradient and the flow-rate response. Figure~\ref{fig:phase lag} shows $\Delta\phi/\pi$ as a function of $\alpha_0$ for different degrees of corrugation $\Delta S$. As $\alpha_0$ increases, the inertia of the fluid prevents it from following the driving pressure instantaneously, and $|\Delta\phi|$ grows. For a given $\alpha_0$, larger corrugations lead to a smaller $|\Delta\phi|$: the reduced flow rate at higher $\Delta S$ implies less fluid inertia, allowing the flow to track the pressure forcing more closely. In the quasi-steady limit $\alpha_0 \to 0$, the pressure and flow are in phase ($\Delta\phi \to 0$) for all geometries, as expected from the Hagen--Poiseuille relation.

\subsection{Wall shear stress} \label{sec:wss}

The wall shear stress (WSS) is a key determinant of particle adhesion and has been linked to plaque development in stenosed arteries~\cite{decuzzi2006,zhou2023,malek1999}. Clinical studies show that low wall shear stress promotes plaque growth, while high wall shear stress is associated with plaque destabilization~\cite{samady2011}. In animal models, wall shear stress is highest within the stenosed segment~\cite{thim2012}. Beyond adhesion, oscillatory wall shear stress patterns in microchannels can propel deformable particles such as red blood cells, with the propulsion speed depending on cell elasticity~\cite{schmidt2022}. The spatial wall shear stress distribution in corrugated channels is therefore relevant not only for stenosis mechanics but also for particle-sorting applications. We focus on the cycle-maximum wall shear stress, $\tau_{\parallel,\max}(x) \equiv \max_t |\tau_\parallel(x,t)|$, and derive the standard hemodynamic metrics (time-averaged wall shear stress and oscillatory shear index) in closed form.

\subsubsection{Spatial distribution}

\begin{figure}[h!]
\centering
\includegraphics[width=\linewidth]{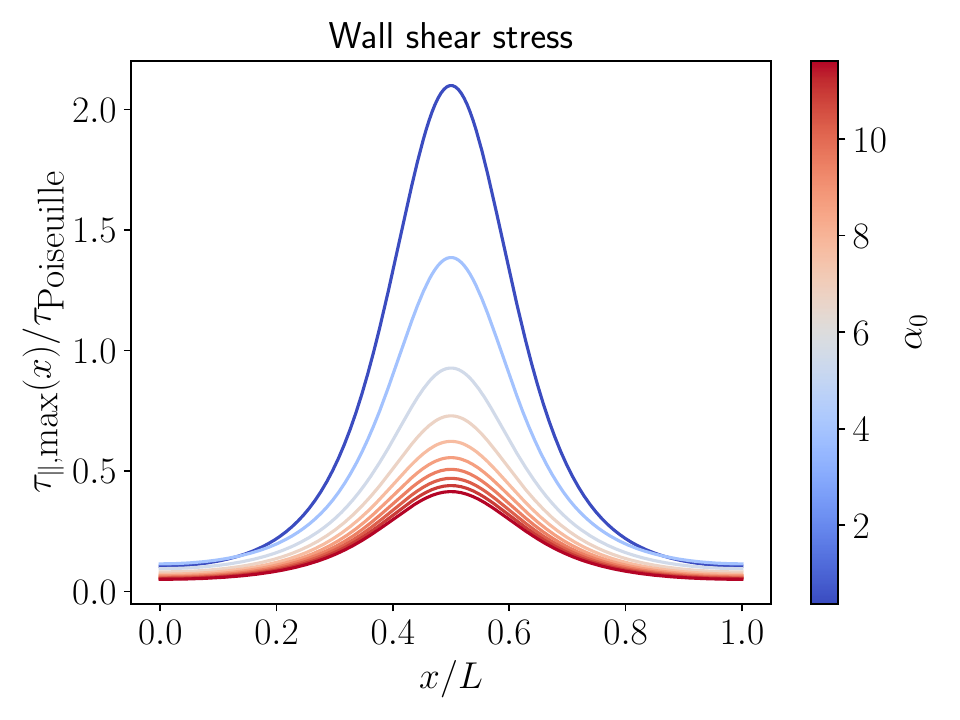}
\caption{Spatial distribution of the cycle-maximum wall shear stress $\tau_{\parallel,\max}(x)/\tau_\text{Poiseuille}$ along a sinusoidally corrugated tube with $\Delta S = 1$ for several $\alpha_0$. The peak always occurs at the bottleneck $x/L = 1/2$.}
\label{fig:wss location}
\end{figure}

Figure~\ref{fig:wss location} shows $\tau_{\parallel,\max}(x)$, normalized by the Poiseuille wall shear stress 
$\tau_\text{Poiseuille} = |\Delta P| R_0 / (2 L)$, along a sinusoidal tube with $\Delta S = 1$ for several $\alpha_0$. For all Womersley numbers, the peak wall shear stress occurs at the bottleneck ($x = L/2$), where the tube radius is minimal and the fluid is forced to accelerate to maintain a uniform flow rate. As $\alpha_0$ increases, the overall magnitude of $\tau_{\parallel,\max}$ decreases.

\subsubsection{Dependence on the degree of corrugation}

\begin{figure}[h!]
\centering
\includegraphics[width=\linewidth]{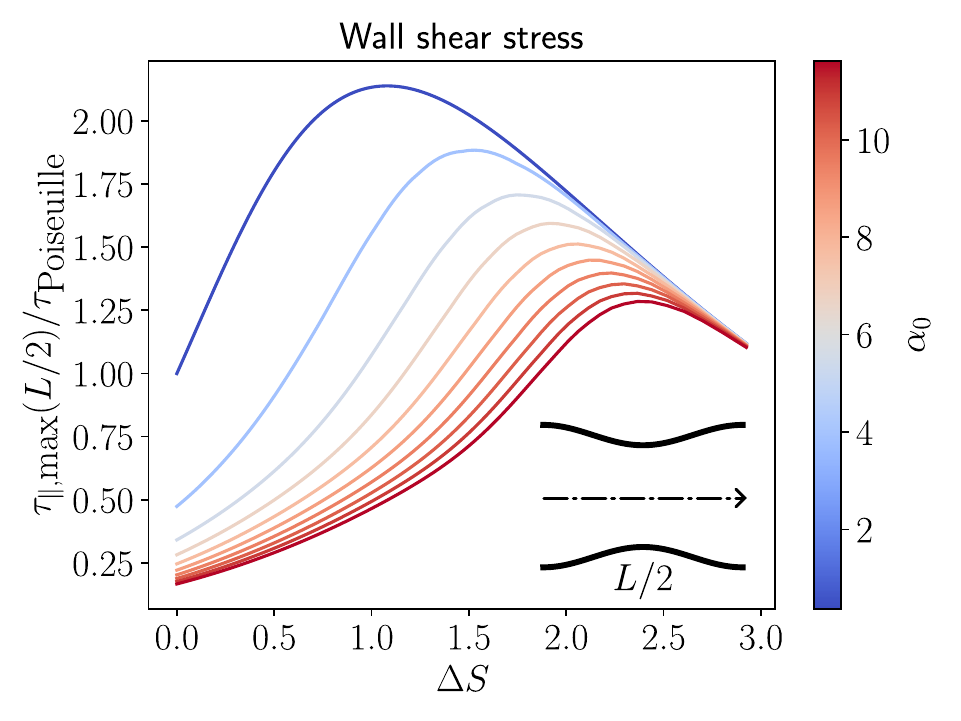}
\caption{Bottleneck wall shear stress $\tau_{\parallel,\max}(L/2)/\tau_\text{Poiseuille}$ vs.\ degree of corrugation $\Delta S$ for a sinusoidal tube at several $\alpha_0$. The nonmonotonic behavior reflects the competition between increasing local velocity and increasing hydraulic resistance.}
\label{fig:wss corrugation}
\end{figure}

Figure~\ref{fig:wss corrugation} reports the bottleneck value $\tau_{\parallel,\max}(L/2)$ as a function of $\Delta S$ for several $\alpha_0$. The dependence on $\Delta S$ is nonmonotonic: the wall shear stress initially increases with corrugation, reaches a maximum at $\Delta S = \Delta S_{\max}$, and subsequently decreases for stronger corrugations. This nonmonotonic behavior reflects a competition between two effects. At moderate corrugation, narrowing the bottleneck increases the local velocity and hence the wall shear stress. At large corrugation, however, the overall hydraulic resistance of the tube grows so strongly that the volumetric flow rate---and with it the velocity at the constriction---diminishes, causing the wall shear stress to decline.

A more quantitative argument proceeds as follows (Appendix~\ref{app:bottleneck_wss}). In the quasi-steady limit, the wall shear stress at the bottleneck scales as $\tau_\parallel \propto Q/R_{\min}^3$, while $Q \propto \langle R^{-4} \rangle^{-1}$. As $R_{\min}$ decreases (increasing $\Delta S$), the factor $R_{\min}^{-3}$ grows, but $Q$ diminishes because the spatial average $\langle R^{-4} \rangle$ is increasingly dominated by the bottleneck contribution $R_{\min}^{-4}$. The product $Q/R_{\min}^3$ therefore peaks at intermediate corrugation. At finite $\alpha_0$, the weaker dependence of $Q$ on $R$ $\left(\propto \langle R^{-2}\rangle^{-1}\right)$ instead of $\langle R^{-4}\rangle^{-1}$ shifts this maximum toward larger $\Delta S$.

The value of $\Delta S_{\max}$ shifts toward larger corrugation as $\alpha_0$ increases, because the plug-like profile at high Womersley number is less sensitive to changes in radius. Additionally, $\tau_{\parallel,\max}(L/2)$ decreases monotonically with increasing $\alpha_0$ at any fixed $\Delta S$; a pulsatile flow thus reduces the peak wall shear stress compared with steady flow. Within the present single-harmonic model, this result suggests that pulsatility attenuates the peak mechanical stress at the constriction relative to a steady flow with the same $|\Delta P|$, which may have implications for the hemodynamic loading of stenotic vessel walls.

\subsubsection{Influence of tube geometry}

To probe the role of the tube shape, we replace the sinusoidal corrugation with a piecewise-linear (triangular) profile in which the radius varies linearly between $R_{\max} = R_0 + R_1$ and $R_{\min} = R_0 - R_1$ over each half-period $L/2$. This geometry shares the same mean radius $R_0$, modulation amplitude $R_1$, period $L$, and degree of corrugation $\Delta S$ as the sinusoidal tube, but distributes the slope $|\partial R/\partial x| = 4R_1/L$ uniformly along the length rather than concentrating it near the inflection points. The lubrication-based analytical framework was validated against LBM simulations for this geometry as well, with relative errors remaining below $3.5\,\%$.

\begin{figure}[h!]
\centering
\includegraphics[width=\linewidth]{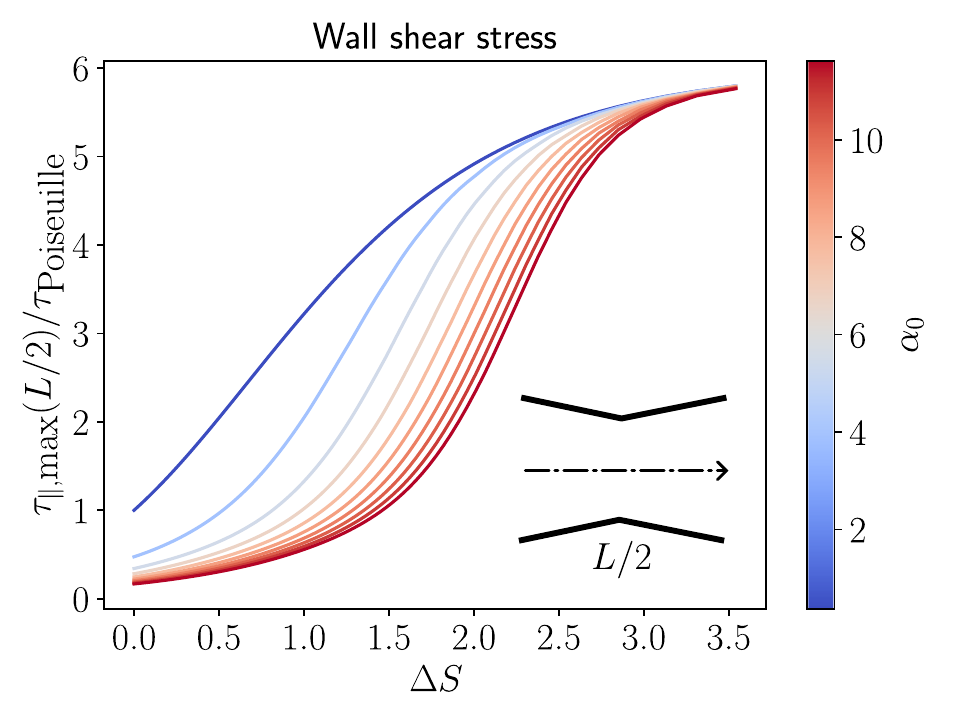}
\caption{Same as Fig.~\ref{fig:wss corrugation} but for a piecewise-linear (triangular) tube. The wall shear stress increases monotonically with $\Delta S$, reaching substantially higher values than in the sinusoidal case.}
\label{fig:wss linear}
\end{figure}

Figure~\ref{fig:wss linear} shows $\tau_{\parallel,\max}(L/2)$ as a function of $\Delta S$ for several $\alpha_0$ for the linearly sloped tube. In contrast to the sinusoidal case, the wall shear stress increases monotonically with $\Delta S$ over the range studied, reaching values more than twice as large. The underlying reason is that, at the same $\Delta S$ (and hence the same $R_{\min}$ and $R_{\max}$), the piecewise-linear tube spends less of its length near the bottleneck radius than the sinusoidal tube does. This yields a smaller value of $\langle R^{-4}\rangle$ and thus a larger flow rate $Q \propto \langle R^{-4}\rangle^{-1}$ at the same bottleneck radius. Since the bottleneck WSS scales as $Q/R_{\min}^3$ (Appendix~\ref{app:bottleneck_wss}), a larger $Q$ at the same $R_{\min}$ produces a larger WSS, and the nonmonotonic turnover is pushed to corrugations beyond the range shown. We note that both geometries satisfy the lubrication condition $|\partial R/\partial x| \ll 1$ for the parameters used here.

\subsubsection{Time-averaged wall shear stress and oscillatory shear index}

The time-averaged wall shear stress (TAWSS) is the standard hemodynamic metric for quantifying the cumulative mechanical loading on the vessel wall~\cite{samady2011,peruzzo2024,he2022}. It is defined as $\text{TAWSS}(x) = 1/T\int_0^T |\tau_\parallel(x,t)|\,dt$, where $T = 2\pi/\omega$. For a single-harmonic pressure forcing, the WSS takes the general form $\tau_\parallel(x,t) = |A(x)|\cos(\omega t + \psi(x))$, where $A(x)$ is the WSS amplitude and $\psi(x)$ is a phase shift. Since the average of $|\cos \theta|$ over one period is $2/\pi$, the TAWSS evaluates to
\begin{equation}
\text{TAWSS}(x) = \frac{2}{\pi}\,\tau_{\parallel,\max}(x) \,,
\label{eq:tawss}
\end{equation}
a universal ratio that is independent of position, corrugation, and Womersley number. All figures showing $\tau_{\parallel,\max}(x)$ can therefore be trivially rescaled to TAWSS by multiplying by $2/\pi$.

The oscillatory shear index~\cite{gao2025}, $\text{OSI}(x) = \frac{1}{2}\bigl(1 - |\int_0^T \tau_\parallel\,dt|/\int_0^T |\tau_\parallel|\,dt\bigr)$, measures WSS directionality and is linked to plaque vulnerability~\cite{thim2012}. For the purely sinusoidal forcing used here, $\int_0^T \tau_\parallel\,dt = 0$ (zero mean), so that
\begin{equation}
\text{OSI}(x) = \frac{1}{2}\,.
\label{eq:osi}
\end{equation}
This is the maximum possible value and follows directly from the zero-mean pressure gradient. For physiologically realistic multi-harmonic pressure waveforms with a nonzero mean component, the OSI would generally be position-dependent and smaller than $1/2$. The present closed-form Eq.~\eqref{eq:wss final} for WSS provides the building blocks for such an analysis via Fourier superposition. Because the lubrication-reduced governing equations and boundary conditions are linear in $u_x$ and $p$, each Fourier harmonic of the pressure forcing can be treated independently. The response to the $n$th harmonic is therefore obtained from Eq.~\eqref{eq:wss final} by replacing $\alpha_0$ with $\alpha_0\sqrt{n}$, and the total velocity and wall shear stress follow by superposition of the harmonic responses. Note, however, that the OSI and $\tau_{\parallel,\max}$ depend nonlinearly on $\tau_\parallel$, through the integrals defining the OSI and through the cycle maximum, respectively. Therefore, they cannot be obtained by superposing individual harmonics.

% ======================================================================
\section{Conclusion}\label{sec:conclusion}

We have derived a closed-form extension of Womersley's solution for oscillatory flow in a rigid circular tube to rigid axisymmetric corrugated tubes within the lubrication approximation. The resulting framework provides explicit expressions for the axial velocity profile, volumetric flow rate, phase lag, and wall shear stress at an arbitrary Womersley number.

Comparison with three-dimensional lattice Boltzmann simulations shows that the theory accurately captures the flow for slowly varying geometries, with relative velocity-profile errors below $3.5\,\%$ for the sinusoidal tube considered here. The analytical solution reveals that pulsatility and geometry interact in a distinctly nonlocal way: the local flow profile depends on the position-dependent Womersley number, becoming plug-like in wide sections and more parabolic near bottlenecks.

Two main consequences follow for transport and wall loading. First, corrugation reduces the cycle-maximum flow rate, but this reduction becomes weaker as the Womersley number increases. In asymptotic terms, the geometric sensitivity crosses over from the quasi-steady Hagen--Poiseuille scaling $Q \propto \langle R^{-4}\rangle^{-1}$ to the weaker high-frequency scaling $Q \propto \langle R^{-2}\rangle^{-1}$. Second, the cycle-maximum wall shear stress is always largest at the bottleneck and decreases with increasing Womersley number. For sinusoidal tubes, the bottleneck wall shear stress depends nonmonotonically on corrugation because local amplification, scaling as $R_{\min}^{-3}$, competes with the reduction in global flow rate caused by the increased hydraulic resistance. For the piecewise-linear geometry considered here, this competition is shifted sufficiently that the wall shear stress increases monotonically over the range studied.

The theory also yields closed-form hemodynamic observables. For single-harmonic forcing, the time-averaged wall shear stress is simply $\mathrm{TAWSS} = (2/\pi)\,\tau_{\parallel,\max}$ and the oscillatory shear index is $\mathrm{OSI} = 1/2$. In addition, the normal viscous traction vanishes at leading order, implying that the viscous loading on the wall is purely tangential within the present approximation.

The present model is restricted to rigid walls, Newtonian fluids, single-frequency forcing, and slowly varying geometries. Within those limits, it provides a transparent analytical baseline for pulsatile transport in constricted tubes and a useful starting point for extensions to compliant walls, non-Newtonian rheology, physiologically realistic multi-harmonic waveforms, and particle or cell transport in corrugated geometries.

\begin{acknowledgments}
We gratefully acknowledge the financial support provided by the Deutsche Forschungsgemeinschaft (DFG) through the research unit FOR2688 `Instabilities, Bifurcations and Migration in Pulsatile Flows' (Project-ID 349558021), project `Impact of inertia on flow-induced particle motion in laminar shear flow' (Project-ID 467503132) and CRC 1411 `Design of Particulate Products' (Project-ID 416229255).
\end{acknowledgments}

\section*{Author Declarations}
\subsection*{Conflict of Interest}
The authors have no conflicts to disclose.

\section*{Author Contributions}
F. Keil: Data curation (lead); Visualization (lead); Methodology (equal); Formal analysis (equal); Software (equal); Validation (equal); Writing – original draft (equal); Writing – review \& editing (supporting).
P. Malgaretti: Conceptualization (equal); Methodology (equal); Formal analysis (equal); Software (equal); Writing – original draft (equal); Writing – review \& editing (equal);  Supervision (equal). 
O. Aouane: Conceptualization (equal); Methodology (equal); Formal analysis (equal); Software (equal); Writing – original draft (equal); Writing – review \& editing (equal);  Supervision (equal).
J. Harting: Conceptualization (equal); Funding acquisition (lead); Resources (lead); Software (equal); Writing – review \& editing (equal).

\section*{Data Availability Statement}
The data that support the findings of this study are openly available at \href{http://doi.org/10.5281/zenodo.20065315}{10.5281/zenodo.20065315}.

\appendix

\section{Auxiliary derivations}
\label{app:auxiliary_derivations}

\subsection{Partial derivatives and reduction of the wall shear stress}
\label{app:partial_derivatives}

At leading order in the lubrication approximation, the axial velocity is
\begin{equation}
u_x(x,r,t) = \Real \left\{ \frac{\ii \mathcal{P}(x)}{\rho\omega} \left[ 1 - \frac{J_0\!\left(\alpha(x)\ii^{3/2}r/R(x)\right)}{J_0\!\left(\alpha(x)\ii^{3/2}\right)} \right] e^{\ii\omega t} \right\}.
\label{eq:app_ux}
\end{equation}
Its radial derivative is
\begin{equation}
\frac{\partial u_x}{\partial r} = \Real \left\{ \frac{\ii \mathcal{P}(x)}{\rho\omega} \frac{J_1\!\left(\alpha(x)\ii^{3/2}r/R(x)\right)\,\alpha(x)\ii^{3/2}}{J_0\!\left(\alpha(x)\ii^{3/2}\right)\,R(x)} \, e^{\ii\omega t} \right\}.
\label{eq:app_dudr}
\end{equation}
A key simplification follows from
\begin{equation}
\alpha(x)\,\ii^{3/2}\,\frac{r}{R(x)} = r\,\ii^{3/2}\sqrt{\frac{\omega}{\nu}},
\label{eq:app_argument}
\end{equation}
which is independent of $x$. Therefore, when differentiating Eq.~\eqref{eq:app_ux} with respect to $x$ at fixed $r$, the numerator Bessel function contributes no explicit $x$-derivative. Moreover, the bracket multiplying $\mathcal{P}(x)$ in Eq.~\eqref{eq:app_ux} vanishes identically at $r = R(x)$, so the terms proportional to $\partial_x \mathcal{P}$ drop out. One obtains
\begin{equation}
\left.\frac{\partial u_x}{\partial x}\right|_{r=R(x)} = - \left.\frac{\partial u_x}{\partial r}\right|_{r=R(x)} \frac{\partial R}{\partial x}.
\label{eq:app_dxux}
\end{equation}

The tangential viscous traction at the wall is
\begin{equation}
\tau_\parallel = \eta\left( -2\frac{\partial u_r}{\partial r}\frac{\partial R}{\partial x} - \frac{\partial u_x}{\partial r} + 2\frac{\partial u_x}{\partial x}\frac{\partial R}{\partial x} \right)_{r=R(x)}.
\label{eq:app_tau_full}
\end{equation}
Axisymmetric incompressibility gives
\begin{equation}
\frac{\partial u_x}{\partial x} + \frac{1}{r}\frac{\partial (r\, u_r)}{\partial r} = 0.
\label{eq:app_incompressibility}
\end{equation}
At the wall, the no-penetration condition $\vec u\cdot\vec n = 0$ gives
\[
u_r\big|_{r=R(x)} = \frac{\partial R}{\partial x}\,u_x\big|_{r=R(x)} = O\!\left(\frac{\partial R}{\partial x}\right).
\]
Thus, at $r=R(x)$,
\[
\frac{1}{r}\frac{\partial (r u_r)}{\partial r}
=
\frac{\partial u_r}{\partial r} + \frac{u_r}{r}
=
\frac{\partial u_r}{\partial r} + O\!\left(\frac{\partial R}{\partial x}\right).
\]
At leading order in $\partial R/\partial x$, Eq.~\eqref{eq:app_incompressibility} therefore reduces to
\begin{equation}
\left.\frac{\partial u_r}{\partial r}\right|_{r=R(x)} = -\left.\frac{\partial u_x}{\partial x}\right|_{r=R(x)}.
\label{eq:app_incompressibility_wall}
\end{equation}
Together with Eq.~\eqref{eq:app_dxux}, this yields
\begin{align}
\left.\frac{\partial u_r}{\partial r}\right|_{r=R(x)} \frac{\partial R}{\partial x} &= O\!\left[\left(\frac{\partial R}{\partial x}\right)^2\right],\\
\left.\frac{\partial u_x}{\partial x}\right|_{r=R(x)} \frac{\partial R}{\partial x} &= O\!\left[\left(\frac{\partial R}{\partial x}\right)^2\right].
\end{align}
Hence, the first and third terms in Eq.~\eqref{eq:app_tau_full} are both $O((\partial_x R)^2)$ and are negligible at leading order. Therefore,
\begin{equation}
\tau_\parallel = -\eta \left.\frac{\partial u_x}{\partial r}\right|_{r=R(x)}.
\label{eq:app_tau_reduced}
\end{equation}
Evaluating Eq.~\eqref{eq:app_dudr} at $r = R(x)$ gives
\begin{equation}
\tau_\parallel = -\eta\, \Real \left\{ \frac{\mathcal{P}(x)\,\ii^{5/2}}{\rho\omega} \sqrt{\frac{\omega}{\nu}}\; \frac{J_1\!\left(\alpha(x)\,\ii^{3/2}\right)}{J_0\!\left(\alpha(x)\,\ii^{3/2}\right)}\; e^{\ii\omega t} \right\},
\label{eq:app_tau_final}
\end{equation}
which is the wall shear stress (WSS) expression used in the main text.

\subsection{Asymptotic limits of the effective conductance}
\label{app:conductance_asymptotics}

To extract the limiting behavior of the flow rate, we introduce the complex effective conductance of a cross-section,
\begin{equation}
\mathcal{D}(x) = R^2(x)\left[ \frac{1}{2}
- \frac{J_1(z)}{z\,J_0(z)} \right],
\label{eq:app_D}
\end{equation}
where $z = \alpha(x)\,\ii^{3/2}$ and $\alpha(x) = R(x)\sqrt{\omega/\nu}$.
In terms of $\mathcal{D}(x)$, the volumetric flow rate reads
\begin{equation}
Q(t) = \Real \left\{ \frac{2\pi\ii\,\Delta P}{\rho\omega}\, e^{\ii\omega t} \left(\int_0^L\frac{dx}{\mathcal{D}(x)}\right)^{\!-1} \right\}.
\label{eq:app_Q}
\end{equation}

We first consider the low-frequency limit $\alpha(x) \ll 1$. In this regime, the Bessel functions can be expanded for small argument as
\begin{equation}
J_0(z) = 1 - \frac{z^2}{4} + O(z^4), \qquad
J_1(z) = \frac{z}{2} - \frac{z^3}{16} + O(z^5).
\label{eq:app_bessel_small}
\end{equation}
Using these expressions, the ratio in Eq.~\eqref{eq:app_D} becomes
\begin{equation}
\begin{aligned}
\frac{J_1(z)}{z\,J_0(z)}
&= \left(\frac{1}{2} - \frac{z^2}{16} + O(z^4)\right)
\!\left(1 + \frac{z^2}{4} + O(z^4)\right) \\
&= \frac{1}{2} + \frac{z^2}{16} + O(z^4),
\end{aligned}
\label{eq:app_ratio_small}
\end{equation}
which implies
\begin{equation}
\frac{1}{2} - \frac{J_1(z)}{z\,J_0(z)}
= -\frac{z^2}{16} + O(z^4).
\label{eq:app_half_minus_ratio}
\end{equation}
Since $z^2 = \alpha(x)^2\,\ii^3 = -\ii\,\alpha(x)^2$, this may be rewritten as
\begin{equation}
\frac{1}{2} - \frac{J_1(z)}{z\,J_0(z)}
= \frac{\ii\,\alpha(x)^2}{16} + O(\alpha(x)^4).
\label{eq:app_small_bracket}
\end{equation}
Substituting this result into Eq.~\eqref{eq:app_D} gives
\begin{equation}
\begin{aligned}
\mathcal{D}(x)
&= R^2(x)\left[\frac{\ii\,\alpha^2(x)}{16}
+ O(\alpha(x)^4)\right] \\
&= \frac{\ii\,\omega\, R^4(x)}{16\nu}
+ O\!\left(\alpha(x)^4\, R^2(x)\right).
\end{aligned}
\label{eq:app_D_small}
\end{equation}

Inserting Eq.~\eqref{eq:app_D_small} into Eq.~\eqref{eq:app_Q}, one finds
\begin{equation}
\left(\int_0^L \frac{dx}{\mathcal{D}(x)}\right)^{\!-1}
\sim \frac{\ii\,\omega\, L}{16\nu}\,
\left\langle R^{-4}\right\rangle^{-1},
\label{eq:app_inverse_integral_small}
\end{equation}
where
\[
\langle f \rangle = \frac{1}{L}\int_0^L f(x)\,dx.
\]
The flow rate, therefore, becomes
\begin{equation}
Q(t) \sim \Real \!\left\{
-\frac{\pi\,\Delta P}{8\eta}\,
\left\langle R^{-4}\right\rangle^{-1}\, e^{\ii\omega t}
\right\},
\label{eq:app_Q_small}
\end{equation}
so that the cycle-maximum flow rate scales as
\begin{equation}
Q_{\max} \propto \left\langle R^{-4}\right\rangle^{-1}.
\label{eq:app_Q_small_scaling}
\end{equation}

In the opposite limit $\alpha(x) \gg 1$, the argument $z = \alpha(x)\,\ii^{3/2}$ is large in magnitude and lies on the ray $\arg z = 3\pi/4$ in the complex plane. Standard large-argument asymptotics for Bessel functions of complex argument imply that, along this ray, $J_0(z)$ and $J_1(z)$ have the same leading exponential and algebraic scaling, namely $O(|z|^{-1/2} e^{|\mathrm{Im}\,z|})$~\cite{watson1944,abramowitz1964}. It follows that $J_1(z)/J_0(z)=O(1)$, so that
\begin{equation}
\frac{J_1(z)}{z\,J_0(z)} = O\!\left(\alpha(x)^{-1}\right).
\label{eq:app_ratio_large}
\end{equation}
The second term in Eq.~\eqref{eq:app_D} is therefore asymptotically small compared with $1/2$, and
\begin{equation}
\mathcal{D}(x) = \frac{R^2(x)}{2} + O\!\left(\frac{R^2(x)}{\alpha(x)}\right).
\label{eq:app_D_large}
\end{equation}
Physically, the plug-like velocity profile at high frequency makes the effective conductance scale with the cross-sectional area $R^2(x)$ rather than with the Poiseuille scaling $R^4(x)$.

Using Eq.~\eqref{eq:app_D_large} in Eq.~\eqref{eq:app_Q}, one finds
\begin{equation}
\left(\int_0^L \frac{dx}{\mathcal{D}(x)}\right)^{\!-1}
\sim \frac{1}{2L}\left\langle R^{-2}\right\rangle^{-1},
\label{eq:app_inverse_integral_large}
\end{equation}
and hence
\begin{equation}
Q_{\max} \propto \left\langle R^{-2}\right\rangle^{-1}.
\label{eq:app_Q_large_scaling}
\end{equation}

The geometric sensitivity therefore crosses over from the quasi-steady Poiseuille scaling $\langle R^{-4}\rangle^{-1}$ to the weaker high-frequency scaling $\langle R^{-2}\rangle^{-1}$.

\subsection{Steady velocity profile in a corrugated tube}\label{app:ratio-v}

In the steady state limit, i.e., $\alpha_0=0$, the axial velocity profile in a corrugated tube reads 
\begin{align}
    u_x(x,r) = \frac{\mathcal{P}(x)}{4\eta}\left(r^2-R^2(x)\right)
\end{align}
where $\mathcal{P}(x)$ is the local pressure gradient. Due to incompressibility, the volumetric fluid rate 
\begin{align}
    Q = 2\pi\int_0^{R(x)} u_x(x,r) r dr =  -\frac{\pi \mathcal{P}(x)}{8\eta}R^4(x)
    \label{app:QQ}
\end{align}
is homogeneous. Hence, the ratio between the maximum axial velocity at the bottleneck, $R_{\min}$, and that at the widest section of the tube, $R_{\max}$, reads 
\begin{align}
     \frac{u_{x,\max}(R_{\min})}{u_{x,\max}(R_{\max})}=\frac{\mathcal{P}(x)|_{R_{\min}}R_{\min}^2}{\mathcal{P}(x)|_{R_{\max}}R_{\max}^2}
\end{align}
which using Eq.~\eqref{app:QQ} reads
\begin{align}
     \frac{u_{x,\max}(R_{\min})}{u_{x,\max}(R_{\max})}=\frac{R_{\max}^2}{R_{\min}^2}
\end{align}

\subsection{Quasi-steady bottleneck wall shear stress}
\label{app:bottleneck_wss}

In the quasi-steady limit ($\alpha(x)\ll 1$), the local wall shear stress is Poiseuille-like,
\begin{equation}
\tau_\parallel(x) = \frac{4\eta\, Q}{\pi\, R^3(x)}\,,
\label{eq:app_tau_poiseuille}
\end{equation}
while the flow rate driven by an imposed pressure drop is
\begin{equation}
Q = \frac{\pi\,|\Delta P|}{8\eta L}\,\left\langle R^{-4}\right\rangle^{-1}.
\label{eq:app_HP}
\end{equation}
Therefore, at the bottleneck $x = x_{\min}$,
\begin{equation}
\tau_\parallel(x_{\min}) = \frac{|\Delta P|}{2L\,R_{\min}^3}\,\left\langle R^{-4}\right\rangle^{-1}.
\label{eq:app_tau_bottleneck}
\end{equation}
Normalizing by the straight-tube Poiseuille value
\begin{equation}
\tau_{\mathrm{Poiseuille}} = \frac{|\Delta P|\, R_0}{2L}\,
\label{eq:app_tau_ref}
\end{equation}
gives
\begin{equation}
\begin{aligned}
\frac{\tau_\parallel(x_{\min})}{\tau_{\mathrm{Poiseuille}}}
&= \frac{1}{R_0\,R_{\min}^3\,\langle R^{-4}\rangle} \\[6pt]
&= \underbrace{\left(\frac{R_0}{R_{\min}}\right)^{\!3}}_{\text{local amplification}}
\;\times\;
\underbrace{\frac{1}{R_0^4\,\langle R^{-4}\rangle}}_{\displaystyle Q/Q_{\mathrm{Poiseuille}}}\,.
\end{aligned}
\label{eq:app_tau_ratio}
\end{equation}
The first factor grows with corrugation, reflecting the local shear amplification produced by a narrower bottleneck. The second factor decreases because the increased hydraulic resistance reduces the total flow rate relative to a straight tube of radius~$R_0$. Their product therefore attains a maximum at an intermediate corrugation, which underlies the nonmonotonic dependence of the bottleneck WSS on~$\Delta S$.

\section*{References}
\bibliography{reference}

\end{document}